\documentclass[sn-mathphys-num]{sn-jnl}

\usepackage{graphicx}%
\usepackage{multirow}%
\usepackage{amsmath,amssymb,amsfonts}%
\usepackage{amsthm}%
\usepackage{mathrsfs}%
\usepackage[title]{appendix}%
\usepackage{xcolor}%
\usepackage{textcomp}%
\usepackage{manyfoot}%
\usepackage{booktabs}%
\usepackage{algorithm}%
\usepackage{algorithmicx}%
\usepackage{algpseudocode}%
\usepackage{listings}%
\usepackage{moreverb}
\usepackage{enumitem}
\usepackage{caption}
\usepackage{lipsum}
\usepackage{soul}

\theoremstyle{thmstyleone}%
\theoremstyle{thmstyletwo}%

\theoremstyle{thmstylethree}%

\begin{document}

\title[Article Title]{Kubernetes in Action: Exploring the Performance of Kubernetes Distributions in the Cloud}


\author[1]{\fnm{Hossein} \sur{Aqasizade}}\email{h.aqasizade@gmail.com}

\author*[1]{\fnm{Ehsan} \sur{Ataie}}\email{ataie@umz.ac.ir}

\author[1]{\fnm{Mostafa} \sur{Bastam}}\email{bastam@umz.ac.ir}

\affil[1]{\orgdiv{Department of Computer Engineering}, \orgname{University of Mazandaran}, \city{Babolsar}, \country{Iran}}




\abstract{
	Kubernetes has emerged as a leading open-source platform for container orchestration, allowing organizations to efficiently manage and deploy containerized applications at scale.
	This paper investigates the performance of four Kubernetes distributions, namely Kubeadm, K3s, MicroK8s,
	and K0s when running OpenFaaS as a containerized service on a cluster of computing nodes on CloudLab. 
	For this purpose, experiments are conducted to examine the performance of two virtualization modes, namely HVM and PV, supported by Xen as the underlying hypervisor. 
	Moreover, two container runtimes that are integrated with Kubernetes, namely Docker, and Containerd, are examined to assess their performance on both disk-intensive and CPU-intensive workloads.
	After determining the appropriate underlying Xen mode and container runtime, the Kubernetes distributions are set up and their performance is measured using various metrics, such as request rate, CPU utilization, and scaling behavior.
	}

\keywords{Kubernetes, Serverless Computing, Performance Evaluation, Benchmarking}



\maketitle

\section{Introduction}\label{sec1}

Cloud computing is transforming a large part of the IT industry by improving the attractiveness of software as a service, as well as changing how hardware is designed and purchased \cite{armbrust2009above}. It offers unlimited storage and processing power, as well as unlimited access to resources \cite{chouat2023adaptive}. 
Virtualization is a crucial technology allowing for efficient hardware resource sharing, encompassing two main types: \textit{virtual machines} (VMs), which emulate complete computer systems \cite{dinesh2024efficient}, and \textit{containers}, lightweight units that share the host operation system kernel for more streamlined deployment and scalability \cite{alfonso2023model, mena2023towards}.

CloudLab \cite{duplyakin2019design}, a cloud computing testbed for research and education, provides an infrastructure for running other platforms, including clusters of virtual machines and containers. 
Xen's default virtualization mode is hardware virtual machine (HVM), which is also supported by CloudLab. 
HVM virtualization allows virtual machines to operate as virtual hardware devices, allowing a wide range of operating systems to be executed concurrently as guest operating systems.
To run paravirtualization (PV) mode on CloudLab, one needs to provide a script to define a new profile. 
With PV, the guest operating system is modified to be aware of its virtualized environment. This mode often results in better performance as it reduces the overhead of virtualization.
As a result of these two approaches, users can tailor the benefits to meet their specific requirements.

\begin{figure}[h]
	\centering
	\includegraphics[width=.8\textwidth]{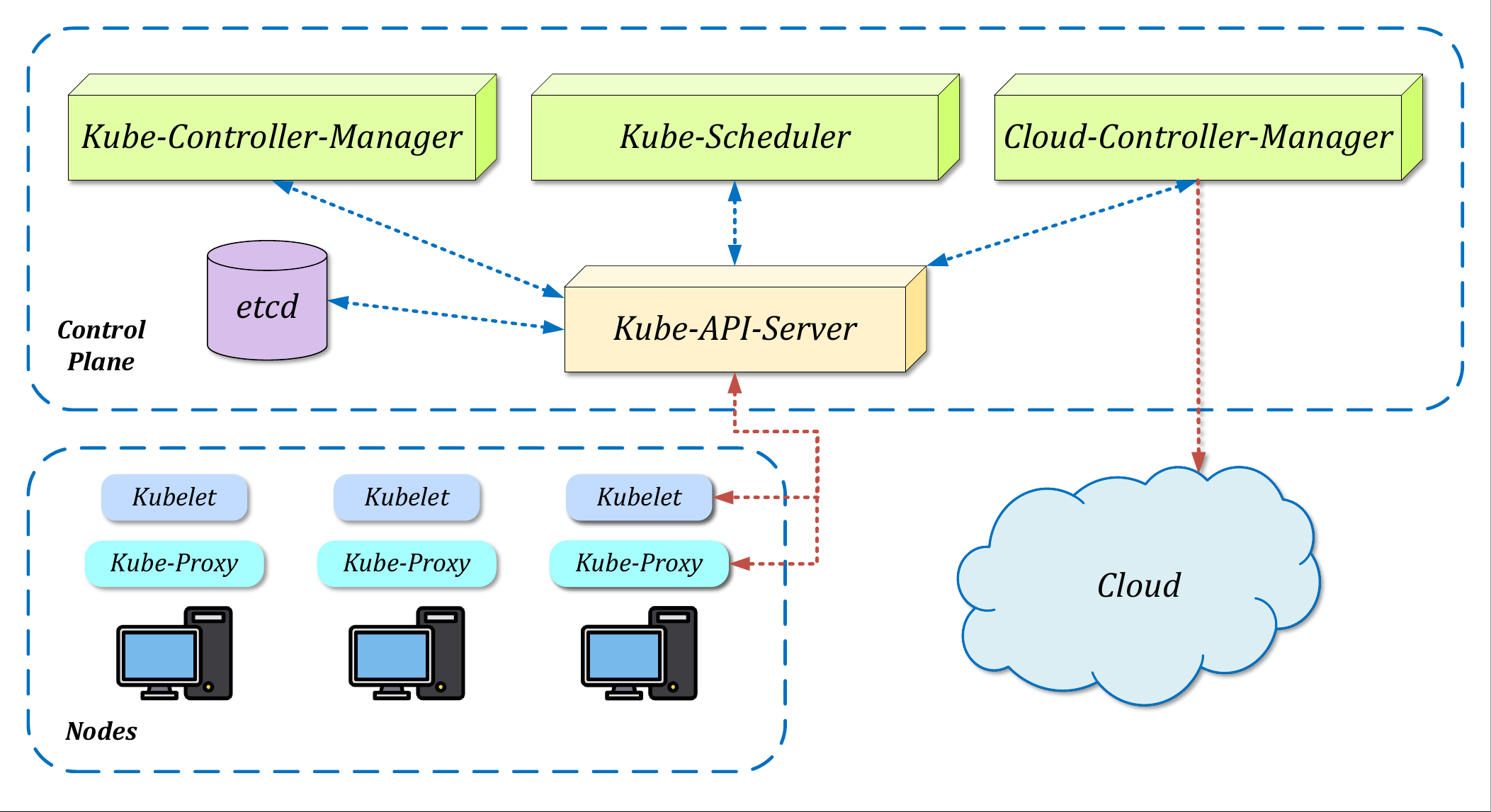}
	\captionsetup{justification=centering}
	\caption{Architecture of Kubernetes}\label{fig2}
\end{figure}

\begin{figure}[h]
	\centering
	\includegraphics[width=.9\textwidth]{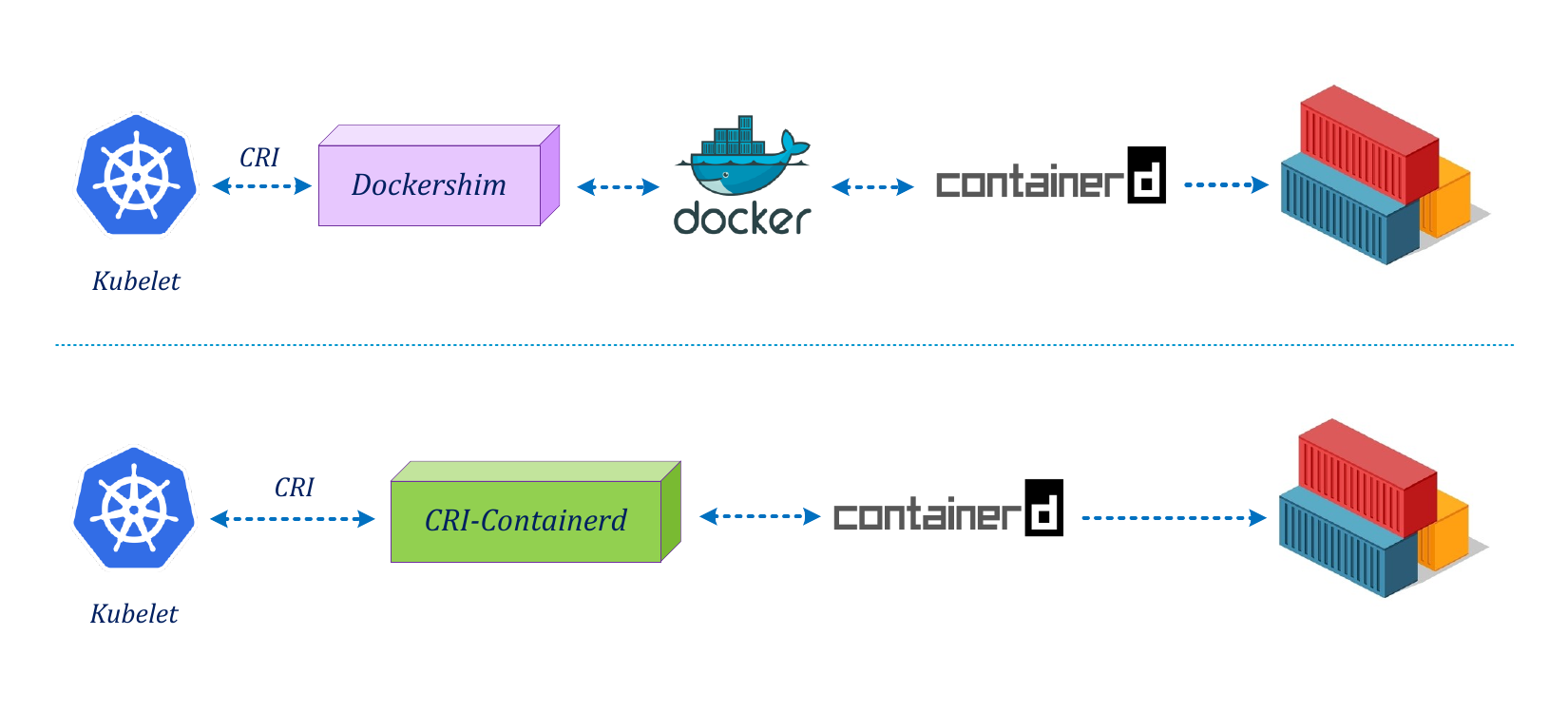}
	\captionsetup{justification=centering}
	\caption{Containerd vs. Docker runtimes}\label{fig3}
\end{figure}

Container orchestration can be provided by Kubernetes {\cite{Kubernetes}}, Docker Swarm {\cite{docker}}, or Apache Mesos {\cite{mesos,rausch2021optimized}}. Kubernetes is widely recognized as a clustering platform to run and manage containerized services such as OpenFaaS. Figure {\ref{fig2}} illustrates the Kubernetes architecture. Kubernetes comes in several distributions, including Kubeadm {\cite{Kubernetes}}, Kubespray {\cite{spray}}, K3s {\cite{K3s}}, Rancher {\cite{rancher}}, MicroK8s {\cite{micro}}, and K0s {\cite{k0s}}.
It includes a container runtime that manages containers' life cycles, including launching, stopping, and monitoring them. The operation of containers can be affected when container runtime use user-space kernels to enforce isolation \cite{app132413329}. Therefore, container runtime performance is important as it ensures the proper and efficient execution of containers, guarantees system performance, and enables a balanced allocation of computing, storage, and network resources. Figure {\ref{fig3}} shows the architectures of Docker and Containerd {\cite{containerd}}, the two most popular container runtimes.

\begin{figure}[h]
	\centering
	\includegraphics[width=.8\textwidth]{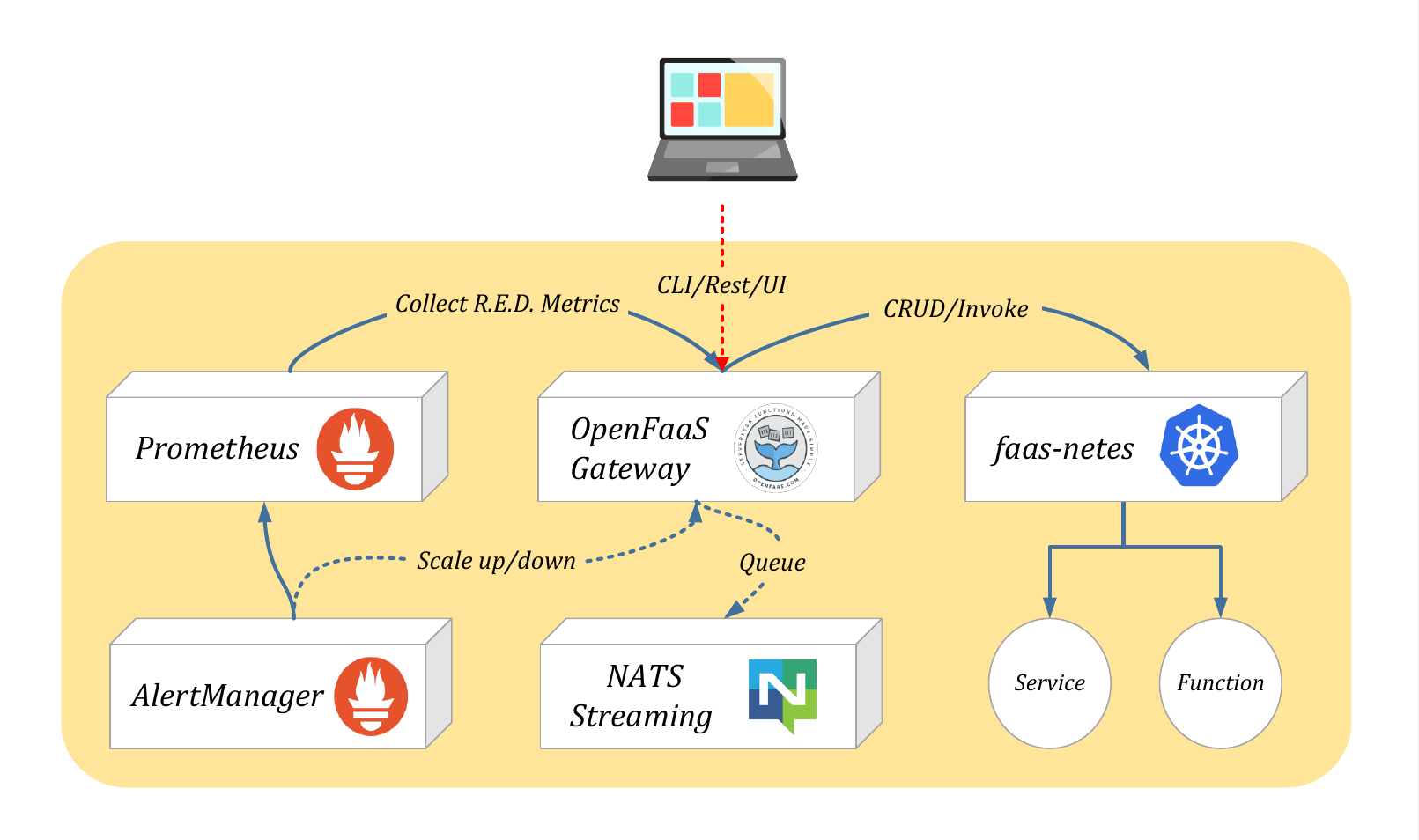}
	\captionsetup{justification=centering}
	\caption{Architecture of OpenFaaS}\label{fig1}
\end{figure}

Function-as-a-Service (FaaS) is a form of serverless computing that abstracts away the underlying infrastructure management, enabling developers to focus on writing and deploying code {\cite{risco2021serverless}}. Such platforms  
are usually integrated with containerization technologies, which let developers encapsulate their dependencies in containers. A FaaS platform manages resource provisioning and function execution automatically based on a developer's function code {\cite{bouizem2023integrating}}. 
There are many popular FaaS platforms, such as Knative {\cite{knative}}, Apache OpenWhisk {\cite{openwhisk}}, Fission {\cite{fission}}, Nuclio {\cite{nuclio}}, but OpenFaaS \cite{OpenFaaS} is one of the most popular due to its github stars (24.3k) \cite{openfaasgit} and its popularity among researchers \cite{10.1007/978-3-030-90539-2_29,9657119,9582259,10305822,10235008,10255010,https://doi.org/10.1002/spe.3277}.
The architecture of OpenFaaS is shown in Figure {\ref{fig1}}.

This study evaluates the performance of four Kubernetes distributions, namely Kubeadm, K3s, MicroK8s, and K0s, from various aspects. 
Since related works have shown that Xen is an efficient hypervisor for virtualization \cite{aqasizade2024experimental, abeni2020using, abeni2019experimental, mavridis2019combining}, this hypervisor is adopted to provide the platform for running Kubernetes. 
At first, we evaluate the performance of both HVM and PV modes to check which mode provides better performance. For this purpose, a new profile is added to CloudLab to provide Xen-PV mode on this testbed. After specifying the appropriate mode, Docker and Containerd, two well-known container runtimes that can be integrated with Kubernetes, are investigated. To this end, Kubernetes has been modified to also use Docker as an alternative to Containerd, so that we can evaluate both runtimes based on disk and CPU performance. To do this, we run Sysbench to compare the runtimes in terms of disk performance when MySQL, as a disk-intensive application, is deployed on Kubernetes. Then, OpenFaaS is run as a compute-intensive application to compare runtime environments in terms of CPU performance. 
When the underlying Xen mode and the container runtime of Kubernetes are adopted, we investigate the performance of the Kubernetes distributions using various metrics.

The remainder of this article is organized as follows:  
Section {\ref{sec2}} provides an overview of the research conducted in this area.
Section {\ref{sec3}} provides the research methodology and the steps of the experiments.
Section {\ref{sec4}} outlines the experimental setup, including hardware and software configurations and the benchmarks. Section {\ref{sec5}} summarizes the experimental results, including comparisons between Xen virtualization modes, container runtimes, and performance behavior of Kubernetes distributions. Finally, Section {\ref{sec6}} draws the final remarks and outlines some future
research directions.

\section{Related Work}\label{sec2}
Recent research indicates that virtualization technologies, containerization, container orchestration, and serverless computing have been topics of considerable interest to many researchers.  
In this section, an investigation of studies on Kubernetes as well as serverless computing platforms is conducted. The literature review aims to be concise, with a focus on research studies relevant to this article.

Koziolek et al. \cite{koziolek2023lightweight} have compared MicroK8s, K3s, K0s, and MicroShift as lightweight Kubernetes distributions. 
The authors found that K3s and K0s marginally showed the highest control plane throughput in stress scenarios, while MicroShift achieved the highest data plane throughput. 
Kjorveziroski et al. \cite{kjorveziroski2022kubernetes} have assessed the performance of three Kubernetes distributions: Kubespray, K3s, and MicroK8s using serverless benchmarks on the OpenFaaS platform. They found that lightweight distributions like K3s and MicroK8s offered better performance than Kubespray, particularly in reducing deployment time and complexity. Despite some performance variations in extreme load tests, K3s and MicroK8s generally performed comparably. 
Costa et al. \cite{costa2023software} have investigated software aging issues within Kubernetes, examining its behavior in a digital twin cloud infrastructure for UAM-ODT systems. Through accelerated lifespan experiments, utilizing Nginx and K3S tools, the authors monitored resource usage and application performance. Notably, they detected memory leaks persisting beyond cluster halts, shedding light on software aging within Kubernetes-based infrastructures.

Mohanty et al. {\cite{mohantypremsankardifrancesco2018}} have examined open-source serverless platforms, including Kubeless, OpenFaaS, Fission, and OpenWhisk. All these platforms, except OpenWhisk, were deployed in a Kubernetes cluster. The authors assessed the features and architectures of each platform and compared them based on response time and success rate under different load conditions. 
They found that the Kubeless framework exhibited more stable performance than Fission and OpenFaaS under various conditions. Palade et al. {\cite{8817155}} have evaluated four open-source serverless platforms, namely Kubeless, OpenWhisk, OpenFaaS, and Knative in an edge computing environment for Internet of Things (IoT) applications. The authors utilized JMeter to assess response time, operational capacity, and success rates of deployed functions under various workloads. As a result, Kubeless demonstrated superior performance in terms of response time and operational capacity to other platforms. 

Balla et al. {\cite{9163456}} have conducted a comprehensive evaluation of the language runtime performances of Python, Node.js, and Golang on platforms such as OpenFaaS, Kubeless, Fission, and Knative. The study explored the impact of supported auto-scaling algorithms on the runtime of the examined functions and proposed solutions to enhance Python runtime performance in OpenFaaS and Kubeless. The research revealed an issue related to session timeout in OpenFaaS, directing all requests to a single function instance and negatively affecting function performance at scale. Additionally, in high concurrent loads, Python execution in Kubeless faced issues due to unresponsiveness to health check messages, which was resolved by adjusting runtimes. Li et al. {\cite{Li_2021}} have analyzed key characteristics of open-source serverless platforms, including Nuclio, OpenFaaS, Knative, and Kubeless, and performance gaps in service exporting and auto-scaling are addressed. Nuclio stood out as having the lowest latency with 99 percentile latency. This is achieved by queuing exclusively within function pods. On the other hand, OpenFaaS and Knative suffer from delays due to multiple components queuing at the same time. Decker et al. {\cite{a15070234}} have revealed an assessment of open-source serverless platforms, including OpenFaaS and Nuclio. They found that Nuclio had 1.5x higher data throughput than OpenFaaS. A key factor contributing to enhanced performance was the integration of user functions with function instances. The study found that existing platforms need improvement for optimal high performance computing (HPC) usage, underscoring the need for broader benchmarking efforts, as well as the limitations of Fission. 
Lee et al. {\cite{8457830}} have evaluated serverless applications for use in research and commercial cloud scenarios, demonstrating their cost-effective deployment in Big Data map-reduce and concurrent image processing. In the study, serverless was demonstrated to be effective in specific applications, such as IoT-driven live migrations in data centers.  

Wen et al. {\cite{https://doi.org/10.1002/smr.2394}} have characterized four mainstream commodity serverless computing platforms (AWS Lambda, Google Cloud Functions, Azure Functions, and Alibaba Cloud Function Compute) through qualitative and quantitative analysis. The researchers have found that too many function instances can lead to poor performance in a single virtual machine. Mavridis et al. {\cite{https://doi.org/10.1002/cpe.6365}} have evaluated security-oriented container runtimes for improved isolation in multi-tenant and serverless cloud scenarios. Additionally, OSv Unikernels and Kubernetes integration was introduced for enhanced isolation. Nabla demonstrated superior performance, while Kata and gVisor showed competitive results. The Kubeless-based serverless framework exhibited longer function start times and reduced performance. Bensalem et al. {\cite{bensalem2023scaling}} have addressed the problem of efficiently scaling and allocating resources for serverless functions in edge networks, especially for delay-sensitive applications, using a reinforcement learning (RL) approach. RL and Deep RL algorithms were compared with monitoring-based heuristics for practical solutions. Simulation results demonstrated that the RL algorithm outperformed monitoring-based algorithms, reducing the total delay of function requests by up to 50\%. Basic RL performed as well as DRL, making it a fast and efficient solution. 

Unlike previous studies, this paper specifically focuses on running Kubernetes on the Xen hypervisor, as an efficient underlying virtualization platform, and harnessing its benefits. Although Kubernetes has replaced Docker with Containerd since version 1.24, another objective of this research is to consider whether Containerd is actually an effective replacement for Docker. Most importantly, this study uses K6 and breakpoint testing to highlight failure points and performance behaviors in different Kubernetes distributions, specifically when OpenFaaS is deployed, which have not been covered in previous studies.

\section{Research Methodology}\label{sec3}
In this study, the experiments are divided into three steps. The specifications and methodologies of each step are discussed in the following sections.

\subsection{Virtualization in CloudLab}
As several studies have found virtualization to be more secure and functional~{\cite{mavridis2019combining, mavridis2017performance, mondesire2019combining, lingayat2018performance}}, virtual machines are considered to be the Kubernetes infrastructure in this study. Further, in our previous work{\cite{aqasizade2024experimental}}, we have found that Xen is an efficient hypervisor for VMs as well as containers running on top of VMs. 
Since CloudLab's default Xen virtualization mode is HVM and lacks PV mode, we have written a Python script to add PV mode, which creates a profile called ``h\_aqasiz"{\cite{maz}}. 
After the PV mode is manually added to CloudLab, IOzone benchmarks are used to compare the performance of the PV and HVM modes. The work is then continued in the virtualization mode, which is empirically proved to be most satisfactory.

\subsection{Runtimes}
In Kubernetes, container runtimes are responsible for load balancing, service creation, and optimum container execution. By default, Containerd is used as the container runtime in recent Kubernetes versions. In this paper, Kubernetes has been modified to also use Docker as an alternative to Containerd, so that we can measure and evaluate the performance of these two runtimes in terms of disk and CPU operation. To do this, we run Sysbench to compare the runtimes in terms of disk performance when MySQL, as a disk-intensive application, is deployed on Kubernetes. Then, OpenFaaS is run as a compute-intensive application to compare runtime environments in terms of processor performance. After determining the winner container runtime, we will continues on that one.

\subsection{Kubernetes distributions}
When the underlying Xen mode and the container runtime of Kubernetes are specified, we investigate four Kubernetes distributions, namely Kubeadm, K3s, MicroK8s, and K0s using various metrics. Kubeadm is a command-line tool for installing and initializing Kubernetes clusters, and it is designed to simplify the boot-up of new clusters on Linux. In contrast, as a lightweight distribution, K3s includes only the necessary components to reduce memory and CPU consumption. It is mainly designed for use in constrained environments such as edge computers and IoT devices. MicroK8s, like K3s, is a lightweight Kubernetes distribution developed by Canonical. It provides a straightforward Kubernetes distribution for development and testing environments. As the newest Kubernetes distribution, Mirantis' K0s can be deployed on both local and cloud networks, simplifying and reducing complexity through Kubernetes' standard design. In order to ensure fairness, all Kubernetes distributions were considered version 1.27.2.

\section{Experimental Setup}\label{sec4}

The purpose of this section is to describe the hardware and software, benchmarks that are used, as well as the execution method and configuration of the applications.

\subsection{Software and Hardware}
The experiment is conducted on C6620 machines in CloudLab, which are equipped with an Intel Xeon E5-2650v2 processor, 64 GB of memory, a 1 TB hard drive, and an X520 PCIe Dual-port 10Gb Ethernet NIC. Ubuntu Server 20.04 {\cite{ubuntu}} is used for all experiments, and the Xen hypervisor is used to virtualize the environment in PV and HVM modes. 
As mentioned earlier, a script was written to create a PV profile in CloudLab for public access. With this profile, one can set the number of nodes and cores, and the value of memory and other parameters.

\begin{figure}[h]
	\centering
	\includegraphics[width=.5\textwidth]{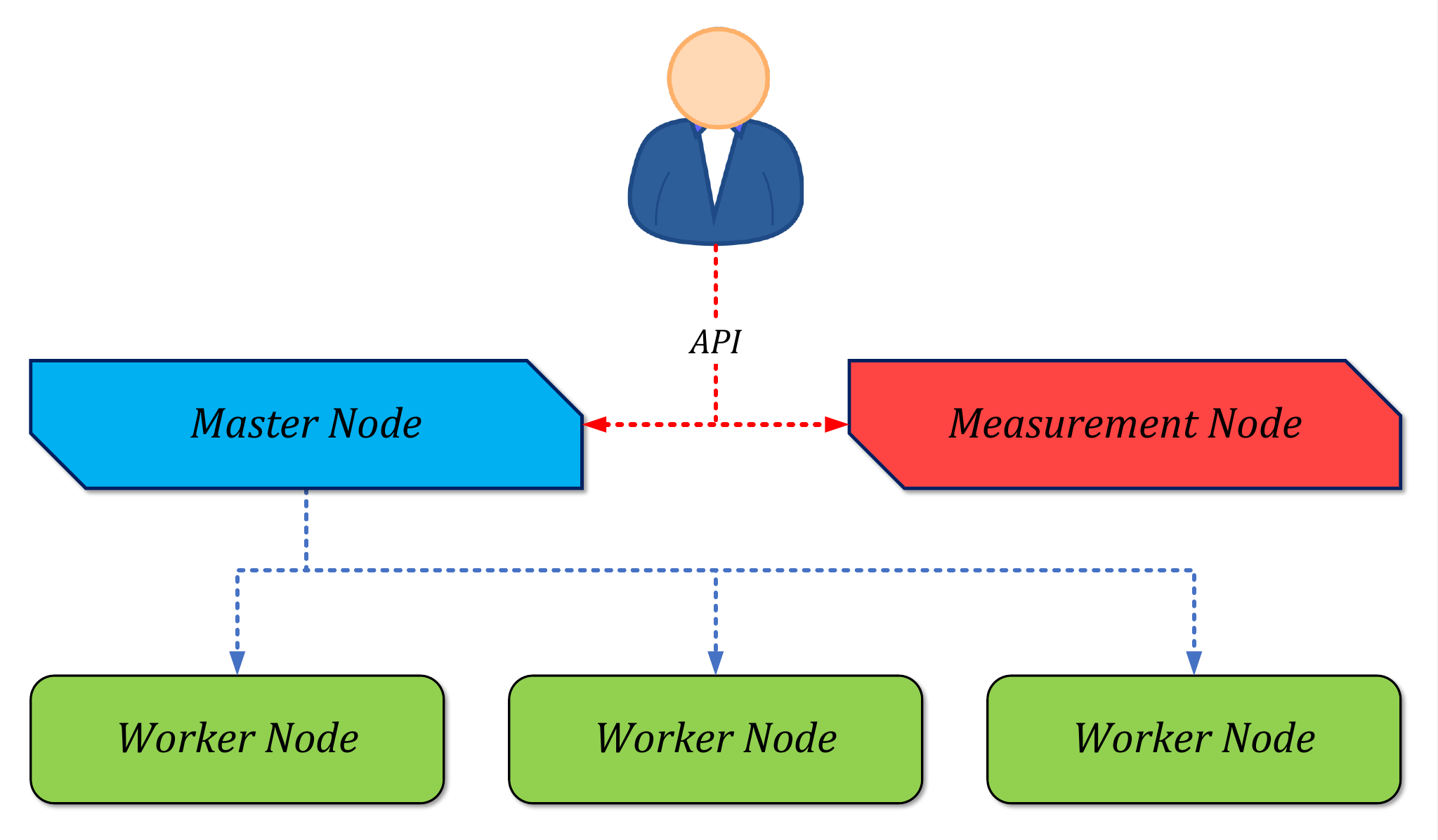}
	\captionsetup{justification=centering}
	\caption{Cluster implementation architecture}\label{fig4}
\end{figure}

Furthermore, we evaluated two popular container runtimes, Containerd and Docker, using the MySQL {\cite{mysql}} service and the OpenFaaS platform in Kubernetes for throughput, CPU usage, and scalability. A serverless service can place considerable load on the processor, so choosing OpenFaaS as an execution service may be a wise choice. Lastly, four Kubernetes distributions were implemented, namely Kubeadm, K3s, MicroK8s, and K0s. All distributions, except Kubeadm, were automatically installed with Calico, but we manually installed it in every distribution to ensure accuracy. The Container Network Interface (CNI) plugin was used by Calico {\cite{calico}}. A master and three worker nodes were used in the study. A separate node was assigned to generate workloads and measure them in order to ensure that the cluster did not interfere with workload generation. 
Figure {\ref{fig4}} illustrates the CloudLab implementation in more detail.

Additionally, we used Prometheus {\cite{prometheus}} and Grafana {\cite{grafana}} to visualize the results, which were installed on the measurement node. In addition, a query was written to obtain values specific to the OpenFaaS and OpenFaaS-Fn namespaces that can be found on our GitHub {\cite{pk8s}}. As a final note, OpenFaaS platform version 0.26.4 has been installed on all Kubernetes distributions using Helm {\cite{helm}} version 3.11.0, and the maximum number of pods was 5 due to OpenFaaS Community Edition. Details of the installation are provided in Table {\ref{tab1}}.

\begin{table}[h]
	\centering
	\caption{Hardware and software specifications in CloudLab}\label{tab1}
	\begin{tabular}{lll}
		\toprule
		\textbf{Criterion} & \textbf{Values}  \\
		\midrule
		Operating System    & Ubuntu Server 20.04 \\
		Number of nodes    & 4 (1 master, 3 workers)  \\
		Processor & Intel\textsuperscript{\textregistered} Xeon\textsuperscript{\textregistered} Silver 4114 \\
		Memory    & 64 GB  \\
		Disk    & 1 TB \\
		Network Card    & Intel X520 PCIe Dual port 10Gb Ethernet NIC  \\
		Kubernetes Version    & 1.27.2  \\
		CNI Plugin    & Calico v3.16  \\
		OpenFaaS Version    & 0.26.4  \\
		Helm Version  & 3.11.0   \\
		\bottomrule
	\end{tabular}
\end{table}

\subsection{Benchmarks}

\begin{figure}[h]
	\centering
	\includegraphics[width=.5\textwidth]{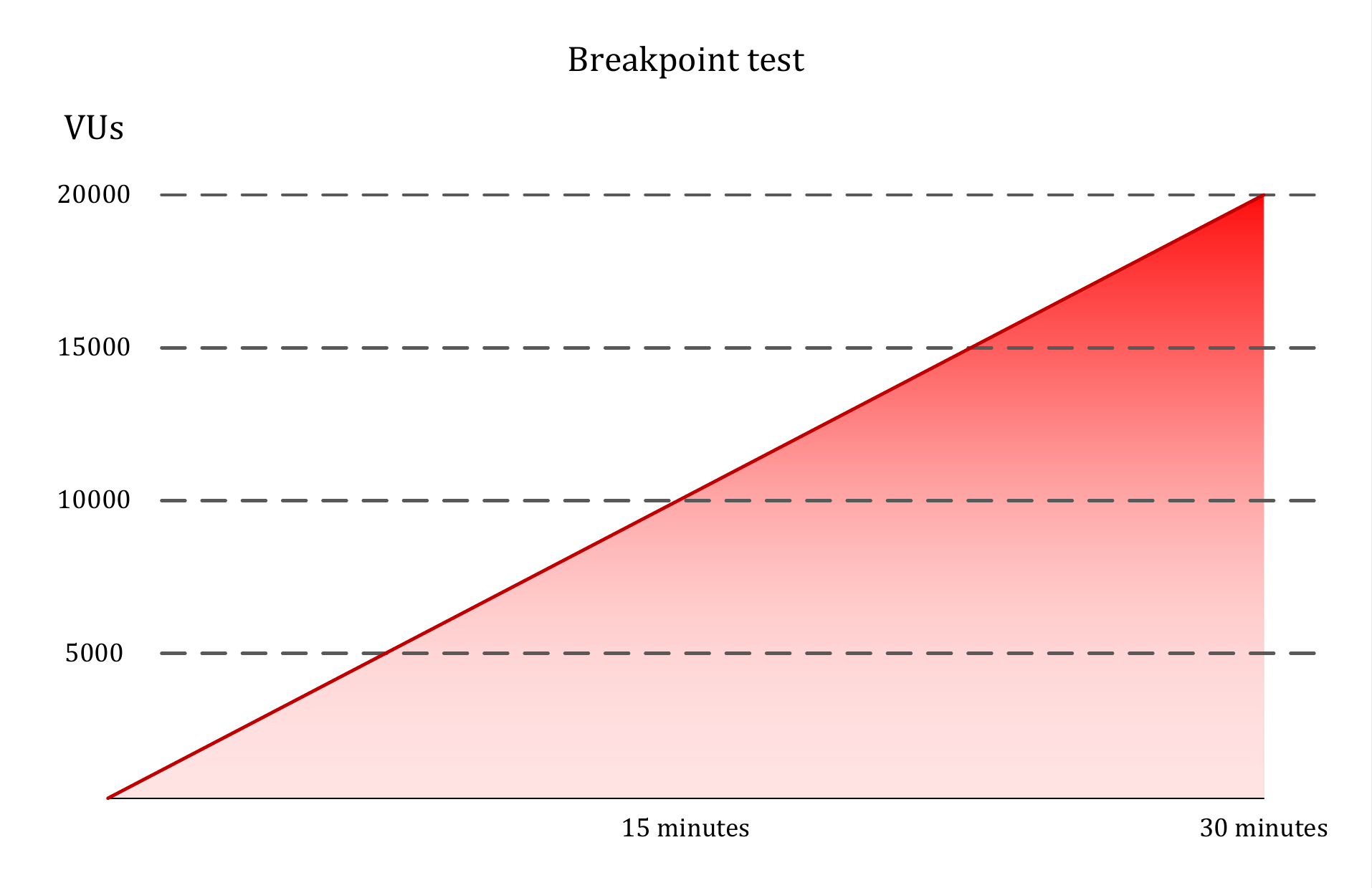}
	\captionsetup{justification=centering}
	\caption{Breakpoint test}\label{fig5}
\end{figure}

There are several tools available for creating workloads on a serverless platform and clustering services, and in this study, the Grafana K6 load testing tool\cite{k6} and Sysbench \cite{sysbench} benchmark were used.   
The breakpoint test from K6 was used to generate workloads for the serverless platform. Breakpoint experiments identify system limitations by progressively increasing the load to an unrealistic level. Understanding how and when a system begins to fail can therefore help to prepare for it. The generated load also allows us to see the function rate as well as other performance behaviors of each Kubernetes distribution. An illustration of the breakpoint test model can be found in Figure \ref{fig5}. Using Sysbench, a script-based multi-threaded benchmark tool, we simulate a heavy load on the MySQL service in Kubernetes by running up to 150 threads simultaneously.

Furthermore, the IOzone {\cite{iozone}} benchmark is also used in this research to compare different modes of virtualization. IOzone is a benchmarking tool for file systems that analyzes the performance of input/output operations by reading and writing to the disk. This study examined a 512 MB file with 2 MB records in the IOzone benchmark to analyze read and write performance. 
As part of OpenFaaS, we implemented a factorial function written in Python and invoked it in order to calculate the factorial of 500, placing pressure on the processor. 
To find the breaking point, concurrency was increased up to 20,000 over a 30-minute period.
According to our observations, it takes almost two minutes for the input queue to become empty after generating a heavy load. 
All measurements were repeated independently and separately 10 times to ensure greater accuracy.
To prevent the experiments from influencing one another, we introduced a 5-minute delay between runs.

\section{Evaluation Results}\label{sec5}
Herein, the results of the experiments conducted for the three mentioned steps of Section {\ref{sec3}} are presented. 
The first section examines disk performance in CloudLab and compares two types of Xen virtualization modes. Next, two different types of container runtimes are examined using MySQL and OpenFaaS. As a final section, the OpenFaaS serverless platform is evaluated from a variety of perspectives when it is run on different Kubernetes distributions.

\subsection{Virtualization in CloudLab}
\begin{figure}[h]
	\centering
	\begin{minipage}[c]{.48\linewidth} 
		\includegraphics[width=\linewidth]{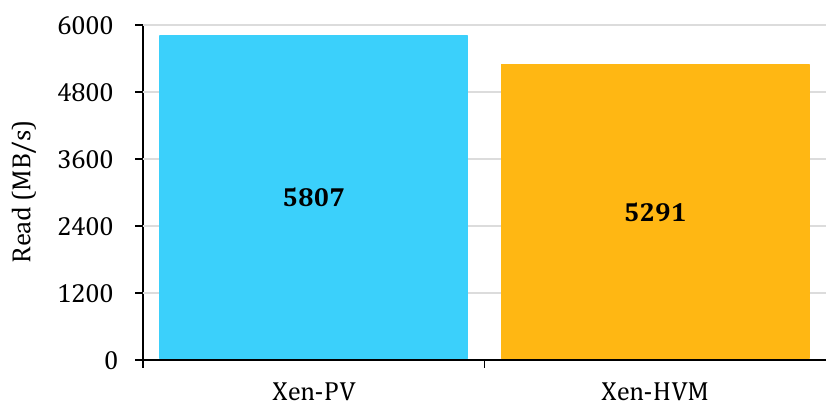}
		\caption{Read performance}\label{fig6}
	\end{minipage}%
	\hfill
	\begin{minipage}[c]{.48\linewidth} 
		\includegraphics[width=\linewidth]{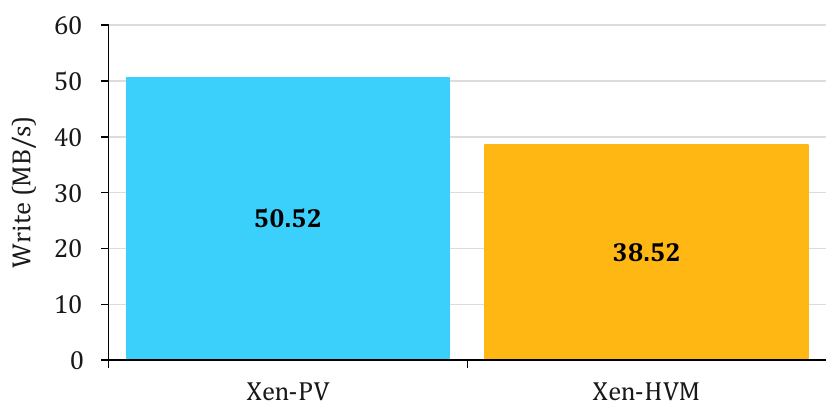}
		\caption{Write performance}\label{fig7}
	\end{minipage}
\end{figure}

This section describes how the underlying virtualization environment is configured on CloudLab servers and how it affects performance. Xen virtual machines are supported by CloudLab, but the default mode is HVM. As mentioned earlier, we have written and executed a Python script in CloudLab to provide Xen-PV mode in that testbed. 
Then, we evaluate the performance of both HVM and PV modes using the IOzone benchmark. The goal is to check which mode provides better disk performance.
Figures {\ref{fig6}} and {\ref{fig7}} show disk read and write performance, respectively, for the two virtualization modes. As can be seen in the figures, PV mode performs significantly better than HVM mode both for reading and writing data. 
To provide additional confidence, we repeated these measurements with a variety of CloudLab settings, including different server locations and machine types. However, the results remain the same.
The results are in line with related work showing that PV offers higher disk performance than HVM. This is because PV drivers allow direct communication between the guest operating system and the hypervisor, thus reducing the overhead associated with emulating hardware in HVM.

\subsection{Runtimes}
This section examines the performance of two popular Kubernetes runtimes, i.e., Docker and Containerd. To this end, the MySQL service is initially executed by both runtimes, and measurements are taken. Then, the OpenFaaS serverless platform is executed by these two runtimes, and the results are examined. For further assurance, the results reported are averages of 10 repetitions.

\subsubsection{MySQL Experiment}
\begin{figure}[h]
	\centering
	\begin{minipage}[c]{.45\linewidth}
		\centering
		\includegraphics[width=\linewidth]{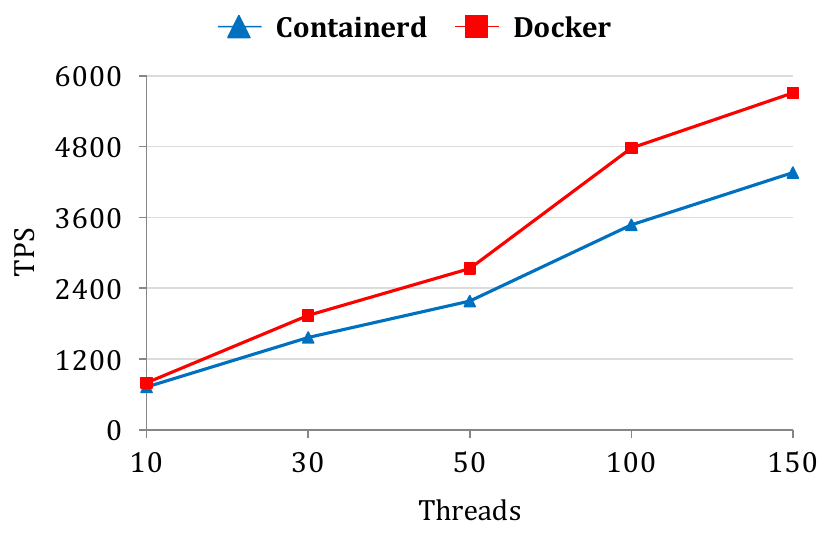}
		\captionsetup{justification=centering}
		\caption{MySQL TPS}\label{fig8}
	\end{minipage}\hspace{0.05\linewidth}
	\begin{minipage}[c]{.45\linewidth}
		\centering
		\includegraphics[width=\linewidth]{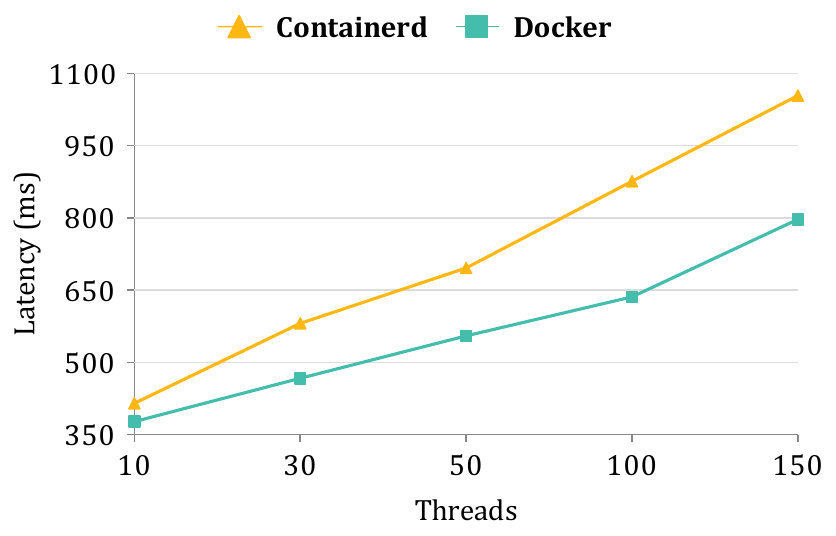}
		\captionsetup{justification=centering}
		\caption{MySQL latency}\label{fig9}
	\end{minipage}
\end{figure}

At first, an experiment is conducted for the MySQL service. The graph in Figure {\ref{fig8}} illustrates the trend in Transactions Per Second (TPS). TPS is not significantly different between the two runtimes at lower thread counts, although Docker has slightly better performance. 
With increasing thread counts, the difference between the two runtimes becomes more apparent, with Docker proving to be the winner under massive loads. 
The most significant difference between Docker and Containerd performance occurs at 100 threads, where Docker shows almost 37\% more TPS than Containerd.
The latency of the two runtimes when MySQL is run is shown in Figure {\ref{fig9}}. As can be seen, the trend is almost the same as Figure {\ref{fig8}} and Docker provides up to 27\% less latency than Containerd. It happens when the number of threads is 100.

\subsubsection{OpenFaaS Experiment}
\begin{figure}[h]
	\centering
	\begin{minipage}[c]{.45\linewidth}
		\centering
		\includegraphics[width=\linewidth]{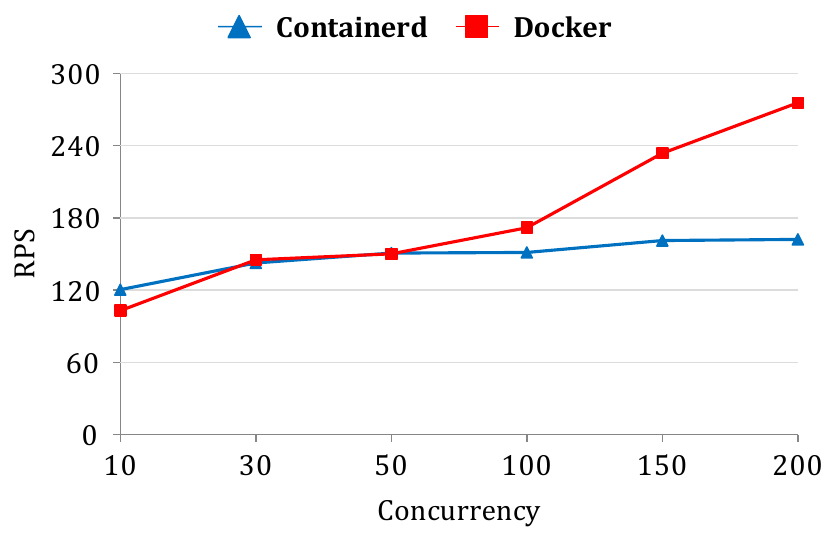}
		\captionsetup{justification=centering}
		\caption{OpenFaaS RPS}\label{fig10}
	\end{minipage}\hspace{0.05\linewidth}
	\begin{minipage}[c]{.45\linewidth}
		\centering
		\includegraphics[width=\linewidth]{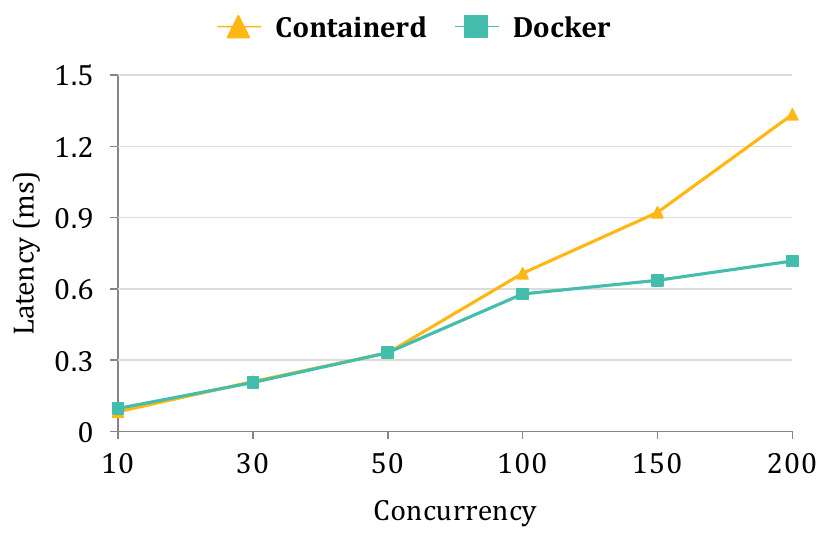}
		\captionsetup{justification=centering}
		\caption{OpenFaaS latency}\label{fig11}
	\end{minipage}
\end{figure}

After examining the Kubernetes runtimes for disk operations using the MySQL service, here, the OpenFaaS serverless platform is used to examine them regarding CPU performance measures.
Figure \ref{fig10} shows the result of assessment in terms of requests per second (RPS) when the concurrency level is changed.
At concurrency 1, Containerd shows better performance than Docker. 
When the concurrency level is 25 and 50, both runtimes perform more or less the same.
As concurrency increases, Docker gradually outperforms Containerd. The maximum difference is at concurrency 200 where Docker can handle almost 70\% more requests per second than Containerd.
Figure {\ref{fig11}} illustrates the latency of the two runtimes under different workloads in a similar manner. It can be seen that Docker shows less latency as concurrency increases. At concurrency 200, OpenFaaS latency, when Docker is adopted, is around 46\% less than when Containerd is chosen.

As a result of the two experiments, it can be concluded that Containerd cannot demonstrate good performance and Docker runtime is a better choice than Containerd for MySQL and OpenFaaS services, particularly under heavy loads.
Docker outperforms Containerd in handling increased concurrencies probably due to its optimized network stack and efficient management of container processes. Docker's architecture, which includes built-in load balancing capabilities, can more effectively distribute incoming requests across containers, thus maintaining higher throughput and lower latency as concurrency levels rise. This could be particularly noticeable in complex applications like MySQL and OpenFaaS, where the overhead of container management and network communication significantly impacts performance.
Therefore, the continuation of this study will be conducted using Docker as the container runtime of Kubernetes.

\subsection{Clustering and OpenFaaS} 

After determining the appropriate underlying Xen mode and container runtime, in this section, various Kubernetes distributions are examined. 
All of the considered distributions, including Kubeadm, K3s, MicroK8s, and K0s, are set up and executed over Xen-PV virtual machines when Docker is set as the Kubernetes container runtime.
For the experimentation, we continue with the OpenFaaS deployment and then invoke the factorial function using breakpoint testing.

\subsubsection{RPS} \label{rps}
\begin{figure}[h]
	\centering
	\includegraphics[width=0.8\linewidth]{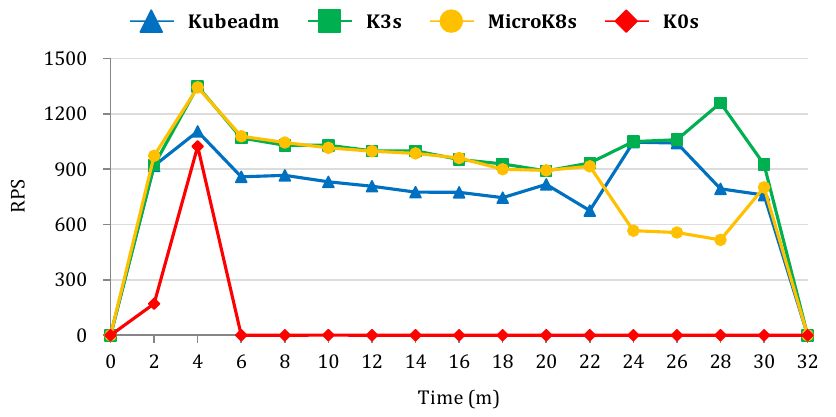}
	\begin{minipage}{0.8\linewidth}
		\captionsetup{justification=centering}
		\caption{Kubernetes distributions function rate}\label{fig12}
	\end{minipage}
\end{figure}

This section examines the request rate for various Kubernetes distributions. Figure {\ref{fig12}} illustrates the performance of distributions in a single frame. 

By applying the workload to various distributions, it appears that all of them grow rapidly. They reach their peak at around minute 4 where K3s and MicroK8s are almost at the same point, taking the lead. 
These two distributions are followed by a significant difference with Kubeadm and then, K0s with a slight difference. As time passes and the rate of input requests increases according to the breakpoint test, queues build up and the function rate gradually decreases. However, it fluctuates within a limited range. In the final minutes, all distributions except K0s grow again.

Regarding K0 behavior, there is a noteworthy point. It turns out that around minute 4, the system experiences a failure, indicated by 503 errors, and the rate of successful responses declines sharply. The system cannot resume normal operation afterwards, so it breaks down. As a result, K0s might perform well under low loads, but under pressure, it fails unlike other distributions.
Performance issues could be related to various factors such as settings, resource constraints, program bottlenecks, or deployment-related issues, and most importantly, to K0s' inherent nature. Overall, K0s does not perform well in the RPS section, and K3s offers the highest performance among the distributions. MicroK8s and Kubeadm are ranked next.

\subsubsection{CPU Consumption} \label{cpu}
\begin{figure}[h]
	\centering
	\includegraphics[width=0.8\linewidth]{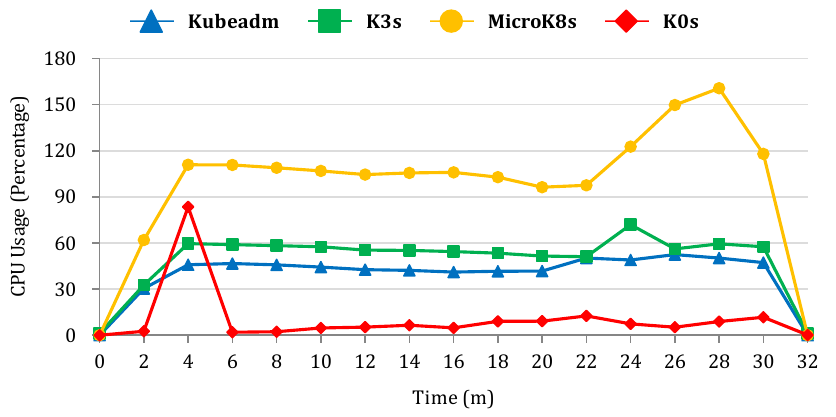}
	\begin{minipage}{0.8\linewidth}
		\captionsetup{justification=centering}
		\caption{Kubernetes distributions CPU usage}\label{fig13}
	\end{minipage}
\end{figure}

The CPU usage for each of the four different distributions is shown in Figure \ref{fig13}. 
Starting from zero, CPU usage gradually increases over time for all distributions. In the 4th minute, when the queue builds up, utilization reaches its local maximum. This is in agreement with the function rate results. 
Afterward, CPU usage remains practically constant with some fluctuations as time passes except for K0s. Between 22nd and 28th minutes the utilization grows up and then, after the load generation stops at 30th minute, it decreases suddenly and becomes zero at minute 32. 
Kubeadm, unlike K3s and MicroK8s, maintains relatively low CPU usage throughout the benchmark, which indicates efficient resource utilization or scalability.

K0s utilization is very low since it drops all requests after minute 6 as explained in Section {\ref{rps}}. MicroK8s has interesting behavior regarding CPU utilization exceeding 100\%, which is the result of  hyper-threading performed by an Intel Xeon Silver 4114 processor with 10 cores and 20 threads. Consequently, it can perform up to 20 tasks at the same time, and each core can increase its frequency up to 3.00 GHz if necessary.  
Very high CPU utilization for MicroK8s may lead to server failure if the CPUs are not powerful enough.
The average CPU usage during the 6th and 22nd minutes, for Kubeadm, K3s, and MicroK8s is around 45\%, 60\%, and 105\%, respectively.

\subsubsection{Scaling}
\begin{figure}[h]
	\centering
	\includegraphics[width=0.8\linewidth]{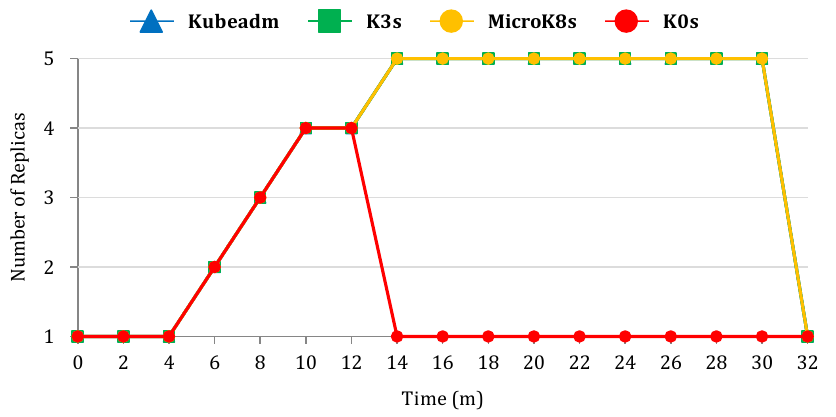}
	\begin{minipage}{0.8\linewidth}
		\captionsetup{justification=centering}
		\caption{Kubernetes distributions replication}\label{fig14}
	\end{minipage}
\end{figure}

Figure \textcolor{blue}{\ref{fig14}} shows a chart illustrating scale changes for different Kubernetes distributions in terms of pod number. All distributions except K0s follow a similar trend, gradually increasing the number of pods to 5, which represents the final value for the community edition of OpenFaaS. However, K0s fails to meet expectations and ultimately decreases its scale from four to just one pod. This result is consistent with previous results explained in Sections {\ref{rps}} and {\ref{cpu}}.

\section{Conclusion}\label{sec6}
In this study, the performances of four popular and widely-used Kubernetes distributions were examined when OpenFaaS was run as a FaaS application. 
At first, The Xen virtualization modes (PV and HVM) in CloudLab environment were assessed, and Xen-PV was selected for use in later phases due to its better performance. After that, the performance of two container runtimes for Kubernetes was evaluated, where Docker was adopted consequently because of up to 37\% higher TPS and 27\% lower latency in MySQL experiments and also up to 70\% higher RPS and 46\% lower latency in OpenFaaS experiments, compared to Containerd.  
Finally, Kubeadm, K3s, MicroK8s, and K0s were investigated when Xen-PV and Docker were set up as the underlying virtualization and container runtime environments, respectively.
It turned out that K03 did not work under high input loads, while the other distributions were able to withstand our breakpoint tests. To sum up, K3s is the winner because it offered a high RPS while maintaining an acceptable level of CPU utilization. Following K3s, MicroK8s demonstrated the highest RPS, but put a heavy load on CPUs. Kubeadm ranks third in terms of RPS, but consumes CPU moderately. 

Machine Learning and Deep Learning techniques are gaining relevance in contemporary computing landscapes, prompting us to explore their potential applications in enhancing Kubernetes and serverless platforms as future work. Several ways can emerge for further investigation and development, such as optimizing resource allocation, identifying bottlenecks and inefficiencies, and integrating anomaly detection. 
Our goal is to unlock machine learning and deep learning's full potential in Kubernetes and serverless platforms through continued research and experimentation.

\bibliography{sn-bibliography}

\end{document}